# A simple strobe to study high-order harmonics and multifrequency oscillations in mechanical resonators

Andres Castellanos-Gomez[1,+,*]

[1] Departamento de Física de la Materia Condensada (C–III). Universidad Autónoma de Madrid, Campus de Cantoblanco, 28049 Madrid, Spain.

[+] Present address: Kavli Institute of Nanoscience, Delft University of Technology, Lorentzweg 1, 2628 CJ Delft, The Netherlands.

E-mail: a.castellanosgomez@tudelft.nl

A simple strobe setup with potential to study higher-order eigenmodes and multifrequency oscillations in micromechanical resonators is described. It requires standard equipment, commonly found in many laboratories, and it can thus be employed for public demonstrations of mechanical resonances. Moreover, the work presented here can be used by undergraduate students and/or teachers to prepare practical works in laboratory courses at physics or engineering universities. The dynamics of a micromachined cantilever is analysed as an example. In fact, using our stroboscopic setup, the first and the second flexural eigenmodes as well as a multifrequency oscillation composed by a superposition of both modes have been successfully filmed with a conventional optical microscope equipped with a digital camera.

## 1. Introduction

The harmonic oscillator is by far one of the most important problems in both classical and modern physics since many different phenomena can be understood within an harmonic oscillator model [1]. Practical works such as the simple pendulum [2] and the spring/mass system or the LC resonant circuit [3] are widely employed to illustrate the physics of oscillatory systems. However, the study of mechanical resonances, which are very interesting for materials/mechanical engineers and physicists, remains challenging. Traditional experimental setups to analyse mechanical resonances rely on the use of very specific components such as holographic microscopes [4] or optical interferometers [5] commonly not present in a teaching laboratory. Recently, stroboscopic methods have been developed to analyse in-plane mechanical resonances using conventional optical microscopes equipped with conventional digital cameras [6–9].

Here, we describe a simple strobe setup with potential to study higher-order eigenmodes and multifrequency oscillations in micromechanical resonators requiring equipment commonly found in many laboratories. We demonstrate the operation of this setup by analysing the dynamics of a micromachined cantilever like the ones commonly used in scanning probe microscopy and biosensing applications.

## 2. Experimental setup: a simple strobe

In this section we describe a simple strobe setup that enables one to study high frequency mechanical resonances (hundreds of KHz). Using the presented strobe the mechanical resonator seems to oscillate in slow motion and this apparent oscillation can thus be filmed with a conventional digital camera.

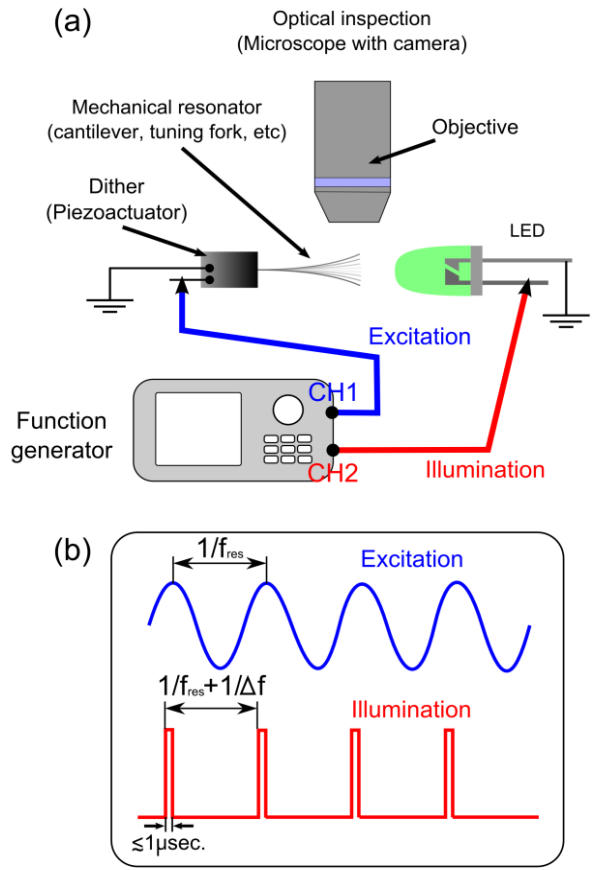

**Figure 1**: (a) Schematic diagram of the stroboscopic setup employed to analyze in-plane oscillations of micromachined mechanical resonators. (b) Illustration of the signals used to drive the mechanical resonator and the illumination.

Figure 1 shows a schematic of the developed strobe setup. A two-channel function generator is employed to both excite the mechanical resonator and to drive the stroboscopic illumination. Two single-channel function generators can be used as well but a small frequency mismatch between them should be expected. One channel is used for exciting the dither piezoelectric attached to the base of the mechanical resonator whereas the other channel is used for exciting a light emitting diode (LED) with short pulses (< 1 μs pulse width). Notice that the pulse width of the stroboscopic illumination can be reduced until a negligible averaging of the resonator motion is achieved. A frequency shift of 1 Hz is typically applied between the mechanical excitation and the illumination to obtain a continuous and smooth phase shift between the excitation and the illumination which yields to an apparent motion of the resonator at 1 Hz. This slow apparent motion of the mechanical resonator can be captured with a conventional digital camera. In order to observe small amplitude mechanical resonances (in the micron scale), the mechanical resonator can be inspected under an optical microscope with a camera attached to the trinocular of the microscope. In our experimental setup, a Nikon Eclipse LV100 optical microscope, with a 50× objective and a Cannon EOS 550D camera attached to the trinocular, has been employed. Note that if a high resolution optical microscope is not available, one can study mechanical resonators with large amplitude resonances, such as those of musical tuning forks, using inexpensive digital microscopes[1] or directly using the zoom lenses of a digital camera (see Supporting Information for a video of the oscillation of a musical tuning fork made directly with a digital camera without microscope).

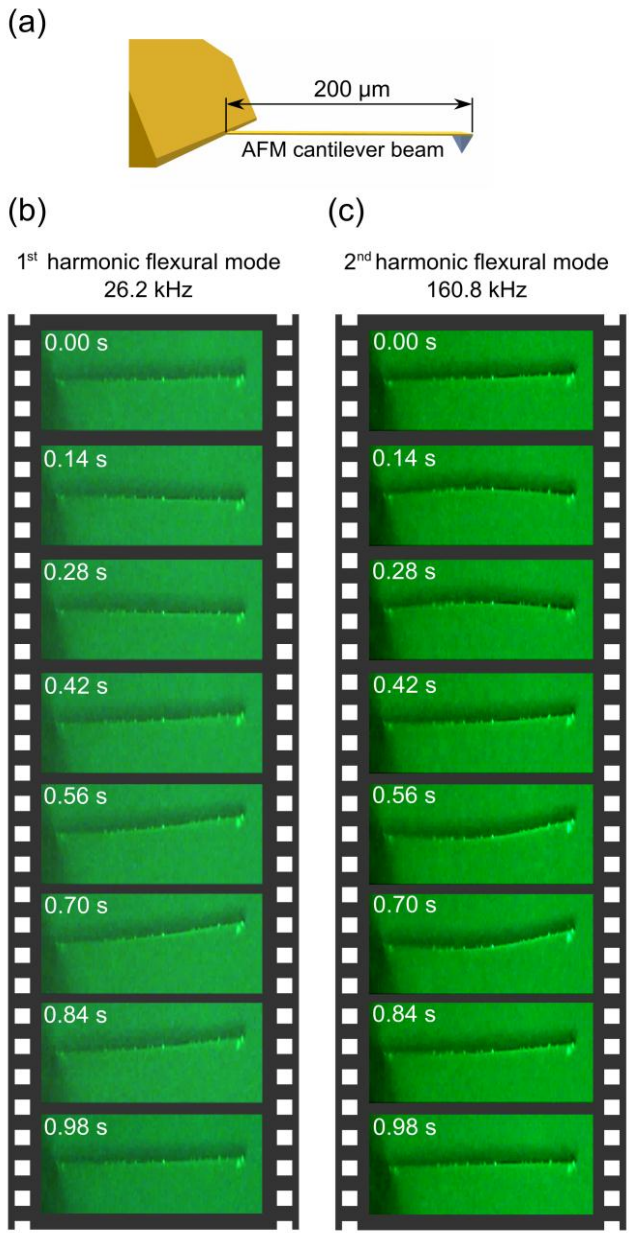

**Figure 2**: (a) Schematic diagram of the micromachined cantilever beam. (b) Filmstrip of the oscillation of the first flexural mode of the cantilever captured with our strobe setup. (c) as (b) for the second flexural mode of the cantilever. The video files can be found in the Supplementary Information linked with this article.

---

[1] www.dino-lite.eu/

We have employed our strobe setup to analyse the resonances of a commercially available micromachined cantilever, used in atomic force microscopes (Figure 2a). Figures 2b and 2c show several frames of the fundamental resonance (at 26.2 KHz) and of the second eigenmode of the cantilever (at 160.8 KHz) filmed with our strobe. The movie files can be found in the supplemental material associated with this manuscript. The frequency ratio between the first and second mode is 6.14 which is very close to the expected value (6.27) calculated from continuum mechanics.

### 3. Experimental setup: a multifrequency strobe

In this section we describe the modification of the simple strobe setup to make possible the study of multifrequency mechanical resonances.

Two different double channel function generators are employed to excite the mechanical resonator and to drive the illumination at two different frequencies (see Figure 3a). One of the electrodes of the dither piezo is driven at the resonance frequency $f_1$ while the other one is driven at the frequency of another resonance $f_2$. The two electrodes of the LED are excited with pulses (< 1 µs of pulse width) at frequencies $f_1 + \Delta f$ and $f_2 + \Delta f \cdot (f_2 / f_1)$. Therefore, a multifrequency resonance can be filmed with apparent frequencies $\Delta f$ and $\Delta f \cdot (f_2 / f_1)$ using a conventional digital camera but maintaining the frequency ratio between the mechanical resonances $f_1$ and $f_2$. The frequencies of the illumination signal have been chosen to maintain the frequency ratio between the mechanical resonances. See the supplemental information associated with this manuscript to see a video of the multifrequency resonance of the micromachined cantilever shown in Figure 2 captured with our multifrequency strobe setup.

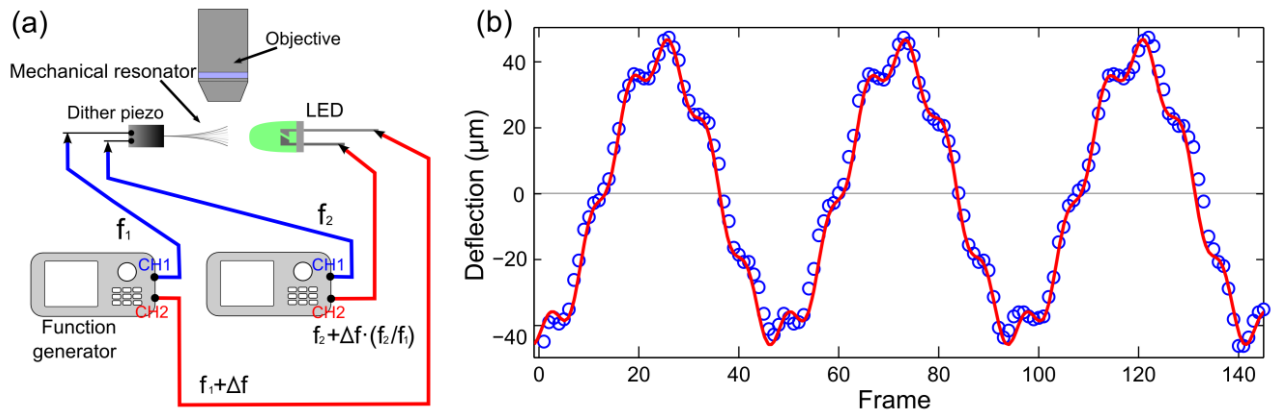

**Figure 3**: (a) Diagram of the stroboscopic experimental setup employed to analyze multifrequency oscillations of micromechanical resonators. (b) Deflection of the cantilever during the multifrequency oscillation. The red line shows the fit of the experimental data to a sum of two sine waves. The video of the multifrequency oscillation of the cantilever can be found in the Supplementary Information linked with this article.

The deflection of the cantilever during the oscillation can be extracted by analysing the captured video. One way to determine the deflection is to extract the position of the free-end of the cantilever frame by frame manually. An alternative procedure relies on the cross-correlation between the first frame and the rest. The position of the maximum correlation indicates the deflection of the cantilever. Figure 3b shows the deflection of the cantilever obtained by analysing the captured multifrequency oscillation with the cross-correlation method. The experimental data has been successfully fitted to the sum of two sine waves with frequencies of 0.13 frames$^{-1}$ (equivalent to1 Hz of apparent frequency and 26.2 KHz of real resonance frequency) and 0.80 frames$^{-1}$ (equivalent to 6.14 Hz of apparent frequency and 160.8 KHz of real resonance frequency) and amplitudes of 42 μm and 5.3 μm respectively. This result demonstrates the capability of this simple strobe to capture multifrequency oscillations even for frequencies up to ~ 200 KHz. We have found that the main limitation to capture multifrequency oscillations with higher frequencies is that the pulse width has to be reduced to avoid averaging over the oscillation of the resonator and thus the intensity of the illumination is accordingly reduced but this can be addressed by employing higher intensity LEDs or a digital camera with a higher gain.

## 4. Conclusions

We have described a simple strobe setup, based on standard laboratory equipment, with capability to study higher-order eigenmodes and multifrequency oscillations in micromechanical resonators. We demonstrated the potential of this setup by studying the dynamics of a micro cantilever. We showed that using this stroboscopic setup it is possible to film the first and the second flexural eigenmodes as well as a multifrequency oscillation, composed by a superposition of both modes, with a conventional digital camera. The simplicity of this stroboscopic method makes it appropriate to be employed by undergraduate students to characterize the oscillation of mechanical resonators. Moreover, its capability to study multifrequency oscillations broadens the possible applications of the proposed setup. As an example of its pedagogical potential, the experimental setup developed in this work was employed by undergraduate students at Universidad Autonoma de Madrid during the course "Tecnicas Experimentales IV" 2009/2010 to study the mechanical resonances of miniature quartz tuning forks (as those present inside wristwatches).

## Acknowledgements

This work was supported by MINECO (Spain) through the program MAT2011-25046.